\documentclass[english,12pt]{article}
\usepackage[cp1251]{inputenc}
\usepackage{babel}
\usepackage{amsmath, amssymb}
\usepackage{axodraw}
\usepackage{epsfig}
\begin{document}
\title{\bf Production of four leptons in electroweak $\gamma\gamma$
interactions} \vspace{0.5cm}
\author{
I. Sotsky  \\ {\it  NC PHEP ,BSU, Minsk, Belarus}}
\date{}
\maketitle
\begin{abstract}
\normalsize Detailed numerical analysis of four charged leptons
photoproduction  in frame of standard theory of electroweak
interaction is presented.  Total and differential cross sections
are obtained and investigated  using  the Monte-Carlo method of
numerical integration. Different energies of initial photons
(60-2000 GeV in c.m.s.) as well as the definite and averaged spin
states of interacting particles  are considered. A set of
kinematical cuts of linear collider experiments is discussed.
\end{abstract}

\section{Introduction}
$ $
 \vspace{-0.5cm}

There are planned to realize high energy experiments on  linear
collider which will have an opportunity to operate in
$\gamma\gamma, \gamma e$ and $e e$ mode  \cite{c1}. This
capability will provide  new advantages in investigation of new
physics, new particles research and detailed study of non abelian
nature of electroweak interaction. However it is necessary to take
into account several important features to realize such
experiments successfully.

 First of all, since produced $W^{\pm}$ and Higgs particles ordinarily decay  within  detector  they  could
be observed via it's decay products only,  for example via several
pairs of leptons in final  state.
  Because of high accuracy and relatively clean environment provided by leptonic
  collider,
   the exact calculation of all backgrounds, namely
 $\gamma\gamma\;\rightarrow 2l$,
$\gamma\gamma\;\rightarrow 2l\; + \;photons$ and
$\gamma\gamma\;\rightarrow 4l\;\;$ processes, are required to
study of such  productions.

 Secondly, since  initial high energy  $\gamma$ quantums are produced by Compton
backscattering of laser ray on  fast electrons, the exact
information about polarization state and energy  spectrum of
obtained photon's beams is necessary to  interpret  the results of
such experiments correctly.
 This data   can be gathered from  measuring of several calibration
 processes, for example:
   $\gamma\gamma\;\rightarrow 2l$,
$\gamma\gamma\;\rightarrow 2l\; + \;photons\;\;$ and
$\gamma\gamma\;\rightarrow 4l$.

So precision investigation of  processes listed behind has great
importance.

The analysis of $\gamma\gamma\;\rightarrow 2l$ and
$\gamma\gamma\;\rightarrow 2l\; + \;photons\;\;$ processes one can
find   in refs. \cite{c2}-\cite{c4}. The detailed theoretical
description of $\gamma\gamma\;\rightarrow 4l$ reaction, a lot of
useful references to the papers devoted to same problem and
expression for squared matrix element  constructed by helicity
amplitude method one can find in ref. \cite{c5}.
 At this paper we perform  numerical
analysis of two  charged leptons pair photoproduction in frame of
Standard Model (SM) of electroweak interaction. A set of
interacting energies, polarization states and  experimental
kinematical cuts are considered and discussed.

 $$ $$ $$ $$
\section{Calculation}
$ $
\vspace{-0.5cm}

There are six topologically different Feynman diagrams  describing
process of $\gamma\gamma\rightarrow 4l$ at the lowest order of SM
(see fig.1).  Whole set of diagrams can be derived on base of
these six ones using C- P- and crossing symmetries.

$$ $$
 \vspace{2cm}
\begin{center}
\begin{picture}(400,100)
\Photon(10,180)(35,167){1}{2} \Photon(10,100)(35,114){1}{2}
\Line(35,167)(85,175) \Line(85,106)(35,114) \Line(45,150)(35,167)
\Line(35,114)(45,130) \Photon(45,150)(45,130){1}{2}
\Line(85,150)(45,150) \Line(45,130)(85,130)
\Text(15,185)[l]{\small $\gamma(k_1)$} \Text(2,115)[l]{\small
$\gamma(k_2)$}\Text(90,177)[l]{\small
$p_1$}\Text(90,150)[l]{\small $p_2$} \Text(90,130)[l]{\small
$p_3$}\Text(90,104)[l]{\small $p_4$} \Text(35,95)[l]{\small
$(1)$}\Text(7,141)[l]{\small
$W,Z,\gamma$}\Vertex(35,167){2}\Vertex(35,114){2}\Vertex(45,150){2}
\Vertex(45,130){2}

\Photon(150,180)(175,167){1}{2} \Photon(150,100)(175,114){1}{2}
\Line(175,167)(225,175) \Line(225,106)(175,114)
\Line(185,140)(175,167) \Line(175,114)(185,140)
\Photon(185,140)(205,140){1}{2} \Line(225,130)(205,140)
\Line(205,140)(225,150) \Text(155,185)[l]{\small $\gamma(k_1)$}
\Text(150,117)[l]{\small $\gamma(k_2)$}\Text(230,177)[l]{\small
$p_1$}\Text(230,150)[l]{\small $p_3$} \Text(230,130)[l]{\small
$p_4$}\Text(230,104)[l]{\small $p_2$} \Text(175,95)[l]{\small
$(2)$}\Text(185,149)[l]{\small
$Z,\gamma$}\Vertex(175,167){2}\Vertex(175,114){2}\Vertex(205,140){2}
\Vertex(185,140){2}

\Photon(290,180)(315,167){1}{2} \Photon(290,100)(315,114){1}{2}
\Line(315,167)(345,159) \Line(365,106)(315,114)
\Line(315,114)(315,167) \Line(345,159)(365,143)
\Photon(345,159)(355,169){1}{2} \Line(365,158)(355,169)
\Line(355,169)(365,180) \Text(290,185)[l]{\small $\gamma(k_1)$}
\Text(290,117)[l]{\small $\gamma(k_2)$}\Text(370,185)[l]{\small
$p_1$}\Text(370,160)[l]{\small $p_2$} \Text(370,145)[l]{\small
$p_3$}\Text(370,104)[l]{\small $p_4$} \Text(315,95)[l]{\small
$(3)$}\Text(317,172)[l]{\small
$W,Z,\gamma$}\Vertex(315,167){2}\Vertex(315,114){2}\Vertex(345,159){2}
\Vertex(355,169){2}

\end{picture}
\end{center}

\vspace{0cm}

\begin{center}
\begin{picture}(400,100)
\Photon(10,180)(50,140){1}{4} \Photon(10,100)(50,140){1}{4}
\Line(65,155)(85,175) \Line(85,106)(65,125)
\Photon(50,140)(65,155){1}{2}\Photon(50,140)(65,125){1}{2}
\Line(85,150)(65,155) \Line(65,125)(85,130)
\Text(15,185)[l]{\small $\gamma(k_1)$} \Text(-3,117)[l]{\small
$\gamma(k_2)$}\Text(90,177)[l]{\small
$p_1$}\Text(90,150)[l]{\small $p_2$} \Text(90,130)[l]{\small
$p_3$}\Text(90,104)[l]{\small $p_4$} \Text(35,95)[l]{\small
$(4)$}\Text(48,155)[l]{\small $W$}\Text(48,125)[l]{\small
$W$}\Vertex(50,140){2}\Vertex(65,155){2}\Vertex(65,125){2}

\Photon(150,180)(175,160){1}{3} \Photon(150,100)(175,120){1}{3}
\Line(200,160)(225,175) \Line(225,106)(200,120)
\Photon(175,120)(175,160){1}{3}
\Photon(175,120)(200,120){1}{2}\Photon(175,160)(200,160){1}{2}
\Line(200,120)(225,130) \Line(225,150)(200,160)
\Text(155,185)[l]{\small $\gamma(k_1)$} \Text(135,114)[l]{\small
$\gamma(k_2)$}\Text(230,177)[l]{\small
$p_1$}\Text(230,150)[l]{\small $p_2$} \Text(230,130)[l]{\small
$p_3$}\Text(230,104)[l]{\small $p_4$} \Text(175,95)[l]{\small
$(5)$}\Text(183,168)[l]{\small $W$}\Text(183,130)[l]{\small
$W$}\Text(183,168)[l]{\small $W$} \Text(160,142)[l]{\small $W$}
\Vertex(175,160){2}\Vertex(175,120){2}\Vertex(200,120){2}
\Vertex(200,160){2}

\Photon(290,180)(315,167){1}{2} \Photon(290,100)(315,114){1}{2}
\Photon(315,167)(350,167){1}{3} \Line(365,106)(315,114)
\Line(315,114)(335,134) \Line(335,134)(365,134)
\Photon(315,167)(335,134){1}{3} \Line(365,158)(350,167)
\Line(350,167)(365,180) \Text(295,185)[l]{\small $\gamma(k_1)$}
\Text(279,110)[l]{\small $\gamma(k_2)$}\Text(370,185)[l]{\small
$p_1$}\Text(370,160)[l]{\small $p_2$} \Text(370,135)[l]{\small
$p_3$}\Text(370,104)[l]{\small $p_4$} \Text(315,95)[l]{\small
$(6)$}\Text(332,175)[l]{\small $W$}\Text(310,145)[l]{\small
$W$}\Vertex(315,167){2}\Vertex(315,114){2}\Vertex(350,167){2}
\Vertex(335,134){2}

\end{picture}
\end{center}
\vspace{-3.5cm}
\begin{center}
{\small Fig.1. Feynman diagrams for process $\gamma\gamma
\rightarrow 4l$.}
\end{center}

The diagrams  containing charged current exchange are excluded
because only processes with four charged leptons in the final
state are considered.

Corresponding cross section has the form:
\begin{eqnarray}\label{c2}
\begin{array}{c}

\sigma = \frac{\displaystyle 1}{\displaystyle
4(k_1k_2)}\int|M|^2d\Gamma, \large
\end{array}
\end{eqnarray}
\noindent where $$d\Gamma
=
\frac{d^3p_1}{(2\pi)^32p_1^0}\frac{d^3p_2}{(2\pi)^32p_2^0}\frac{d^3p_3}{(2\pi)^32p_3^0}\frac{d^3p_4}{(2\pi)^32p_4^0}(2\pi)^4\delta(k_1+k_2-p_1-p_2-p_3-p_4)$$

\noindent is phase space element, $p_{i}$- four momentum of final
lepton and $k_{i}$- four momentum of initial photon and
$(k_1,k_2)$- scalar product of two four vectors.

The construction of squared matrix element  is realized both using
helicity amplitude method \cite{c6}-\cite{c9} as well as precision
covariant one (see, for example, refs \cite{c10},\cite{c11}).
Helicity amplitude method is used to obtain cross sections  in
massless limit at each possible  polarization state configuration
of interacting particles. Due to compact form of final expression
it is allowed to perform  numerical integration with high enough
accuracy for a short interval of computing time. The explicit
matrix element expression constructed by using helicity amplitude
method one can find in ref. \cite{c5}.

The covariant method of precision calculation allows to write
squared matrix element without any approximation. It is applied to
estimate  error of massless approximation and to investigate heavy
leptons production processes. The last method was used for
calculation of cross section in case of unpolarized particles
interaction.

The investigation of differential  and  total cross sections  is
realized by using of the Monte-Carlo method of numerical
integration.

\section{Results and Conclusion}
$ $
 We have calculated the differential and total cross
sections of $\gamma\gamma\;\rightarrow 4l$ process in frame of SM
at set of initial particles energies, polarization states and
kinematical cuts.

The values of total cross sections at different kinematical
conditions  are presented in the Table. The differential cross
section dependence on cosine of polar angle (angle between final
lepton and initial photon) is shown at figs. 2-9. The case of
averaged polarization state of interacting particles and fixed one
are considered.

Presented results demonstrate the  expected dependence of cross
section on cosine of polar angle: cross section is almost equal to
zero at the center  of kinematical region (the kinematical region
where final particles have large value of $p_{\bot}$) and
extremely increases if one  closed  to the borders (the
kinematical region with small $p_{\bot}$) (see figs. 2-9). The
total cross section also strongly increases with decreasing of
interacting photons energy (see the Table and figs. 14,15), owing
to its expression   contains $\frac{\displaystyle 1}{\displaystyle
k_1k_2}$ factor, where $k_1$ and $k_2$ are four momenta of initial
photons.

The polarized assymetry of differential cross section  (figs. 4,5)
occurs because of  the probability of  lepton  scattering into
direction closed to momentum of initial photon  is larger in case
of different signs of photon polarization  and scattering lepton
helicity  then in case of the same signs.

As one can see at figs. 6,7 the differential cross section
dependencies are symmetric  since the initial  photons have the
similar polarization state, but the cross section value  is larger
when the sign of lepton helicity is opposite to polarization signs
of interacting photons. The total cross section of lepton
photoproduction with fixed polarization states  is larger in case
of different polarizations of initial photos then in case of
similar ones (see figs. 8,9).

At figs. 10-13  the differential cross section dependencies on a
set of  kinematics variables are presented. Peaks which one can
see at fig.10 are  so colled collinear peaks. This peaks appear
when several particles propagate into very closed direction. Cross
section incidence near borders  of  kinimatical region is caused
by application of kinematical  cuts applying. Peak at fig. 11 is
well noticeable so called Z-peak,  which is appeared if mass of
any final pair is closed to mass of Z-boson. At fig. 12 one can
see that cross section extremely increases when energies of two
final electron positron pairs are closed to each other. Fig. 13
demonstrates that cross section has the largest value when any
final particle has the minimal (or the maximal) admissible value
of  energy.

The comparison of results obtained by the helicity amplitude
method and the precision covariant one indicates that electron and
positron mass contributions are vanishingly small under
investigated kinematical condition: energy of initial photons
beams is $60 \sim 2000$ GeV, polar angle is larger then $7^{o},$
and angle between any two final particles is larger then $3^{o}$ .

Besides that, as one can see in the Table, the distinguish between
cross sections of processes  electrons and  muons photoproduction
becomes less then statistical error of Monte-Carlo method  when
energy of interacting particles equals to 0.5 TeV in c.m.s. So if
the energy of initial photons is equal to or higher then 0.5 TeV
in c.m.s  the muon mass contribution could be neglected in frame
of precision of planned experiments. However, it is imposible to
neglect $\tau$ lepton mass contribution anywhere at investigated
kinematical field (see the Table).

 The relative error of obtained results ($\sim$
0.9\%) is less then expected experimental error   in future high
energy experiments on linear collider \cite{c1}.

We have obtained the differential and total cross sections in
frame of  precision covariant method for unpolarized interacting
particles only, so it is impossible to use them for investigation
of polarized leptons production processes. However, it was
discovered that  one can perform spin effects analyzing in $\gamma
\gamma \rightarrow \mu^+\mu^-\mu^+\mu^-$ and $\gamma \gamma
\rightarrow e^+e^-\mu^+\mu^-$ reactions with enough precision
applying  matrix elements expression constructed in frame of
helicity amplitude method, if interacting energy is larger then
0.5 TeV.

The differential cross section of $\tau$ leptons production have
the typical as $e$- and $\mu$- production  form but its value is
significantly smaller (see, for example, fig. 2). At fig. 14 the
interesting feature of total cross section dependence on energy of
interacting particles is presented. The value of  total cross
section of two $\tau$ leptons pairs photoproduction is less then
cross section of processes with two different leptons pairs
photoproduction  ($\gamma \gamma \rightarrow ee\mu\mu$, $\gamma
\gamma \rightarrow ee\tau\tau$ and $\gamma \gamma \rightarrow
\mu\mu\tau\tau$) at energy up to 150 GeV. The electrons and muons
cross sections, for example, have the different behavior (see
figs. 14,15).

\begin{center}

The Table. The  total cross sections dependencies  on energy of
initial particles. Here the following notation $(\alpha,\beta)$ is
used to describe kinematical cuts, where $\alpha$ is the minimal
angle between  any two final particles, $\beta$ -- the  minimal
polar angle. Minimal admissible energy of any final particle is
$1$ GeV.
\end{center}

 \small{\begin{center}

\begin{tabular}{|c|l|l|l|}
 \hline
 \multicolumn{1}{|c|}{cut} & \multicolumn{3}{|c|}{$(3^o,7^o)$} \\ \hline
\multicolumn{1}{|c|}{energy (Gev)}
&\multicolumn{1}{|c|}{$e^+e^-e^+e^-$}&\multicolumn{1}{|c|}{$\mu^+\mu^-\mu^+\mu^-$}&\multicolumn{1}{|c|}{$\tau^+\tau^-\tau^+\tau^-$}
\\ \hline
 $60$&$1254.36\pm 6.73$&$1210.99\pm 8.90$&$269.98\pm 2.82$
 \\ \hline
 $120$&$379.86\pm 1.87$&$370.37\pm 2.48$&$179.01\pm 1.52$ \\ \hline
 $200$ &$154.94\pm 0.81$&$152.87\pm 1.25$ &$93.85\pm 0.63$ \\ \hline
 $300$&$76.41\pm 0.47$&$75.36\pm 0.44$ &$52.45\pm 0.35$ \\ \hline
 $400$&$46.97\pm 0.30$&$46.67\pm 0.35$&$34.46\pm 0.30$ \\ \hline
 $500$&$32.15\pm 0.15$&$32.14\pm 0.27$&$24.15\pm 0.15$ \\ \hline
 $1000$&$9.90\pm 0.07$&$9.97\pm 0.12$&$8.35\pm 0.08$ \\ \hline
 $1500$&$5.05\pm 0.04$&$5.07\pm 0.11$&$4.31\pm 0.03$ \\ \hline
 $2000$&$3.04\pm 0.03$&$3.03\pm 0.03$&$2.66\pm 0.02$\\ \hline

\multicolumn{1}{|c|}{}
&\multicolumn{1}{|c|}{$e^+e^-\mu^+\mu^-$}&\multicolumn{1}{|c|}{$e^+e^-\tau^+\tau^-$}&\multicolumn{1}{|c|}{$\mu^+\mu^-\tau^+\tau^-$}
\\ \hline
 $60$&$630.20\pm 5.49 $&$354.83\pm 2.29$&$341.28\pm1.42$ \\ \hline
 $120$&$189.41\pm 1.55 $&$139.00\pm 1.09$&$133.51\pm 0.90$ \\ \hline
 $200$&$78.12\pm 0.54 $&$62.80\pm 0.46$&$62.31\pm 0.52$  \\ \hline
 $300$&$38.77\pm 0.23 $&$32.52\pm 0.23$&$32.40\pm 0.27$ \\ \hline
 $400$&$23.95\pm 0.17 $&$20.52\pm 0.16$&$20.49\pm 0.12$ \\ \hline
 $500$&$16.43\pm 0.14 $&$14.18\pm 0.12$&$14.15\pm 0.13$ \\ \hline
 $1000$&$5.03\pm 0.04 $&$4.65\pm 0.04$ &$4.56\pm 0.03$  \\ \hline
 $1500$&$2.55\pm 0.01 $&$2.33\pm 0.02$  &$2.30\pm 0.01$   \\ \hline
 $2000$& $1.52\pm 0.01 $&$1.42\pm 0.01$  & $1.40\pm 0.01$ \\ \hline

\end{tabular}

\end{center}
}

 \vspace{-1cm}

\begin{center}
\begin{figure}[ht!]
\leavevmode \centering
\includegraphics[width=8.5cm, height=7.5cm, angle=0]{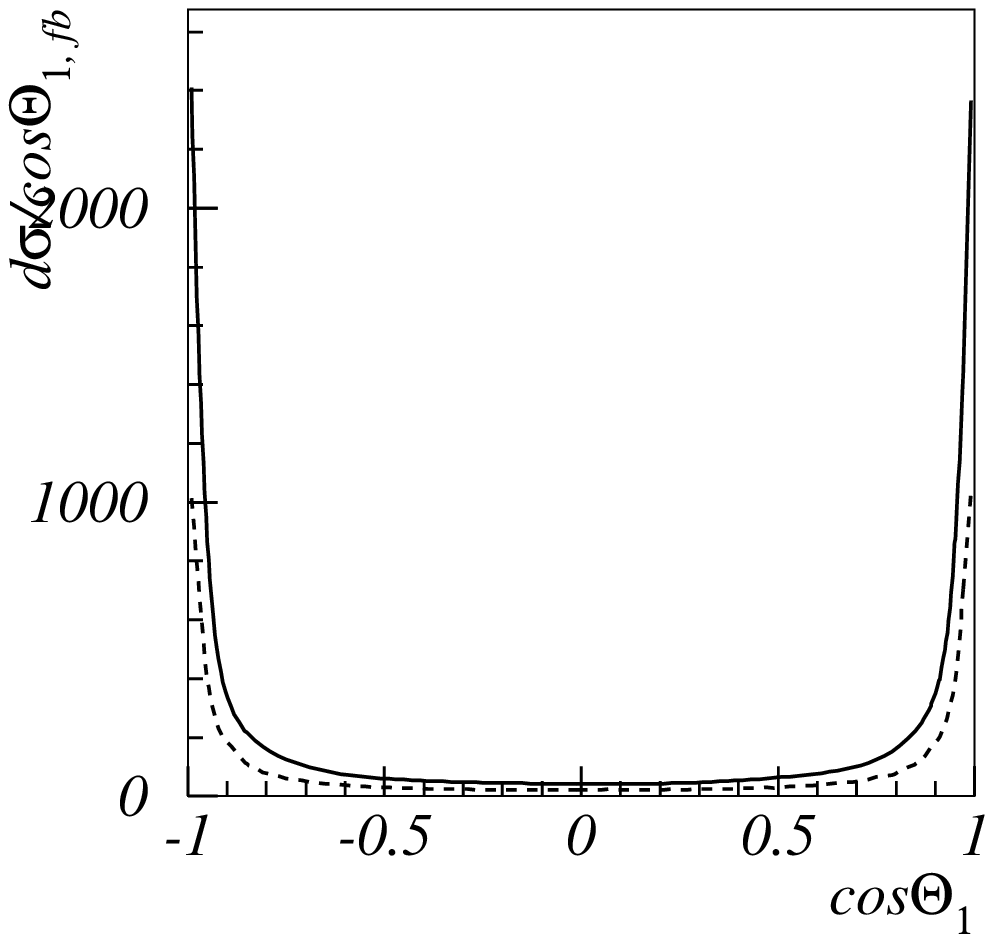}

 \vspace{-15pt}

 { Fig.2. The differential cross section dependence of $\gamma\gamma \rightarrow
e^+e^-e^+e^-$ (solid line) and $\gamma\gamma \rightarrow
\tau^+\tau^-\tau^+\tau^-$ (dashed line)  processes on cosine of
polar angle at averaged polarization state of interacting
particles. The energy of $\gamma\gamma-$ beam is $ 120$ GeV in
c.m.s. }

\end{figure}
\end{center}
\vspace{-1cm}

\begin{center}
\begin{figure}[ht!]
\leavevmode \centering
\includegraphics[width=8.5cm, height=7.5cm, angle=0]{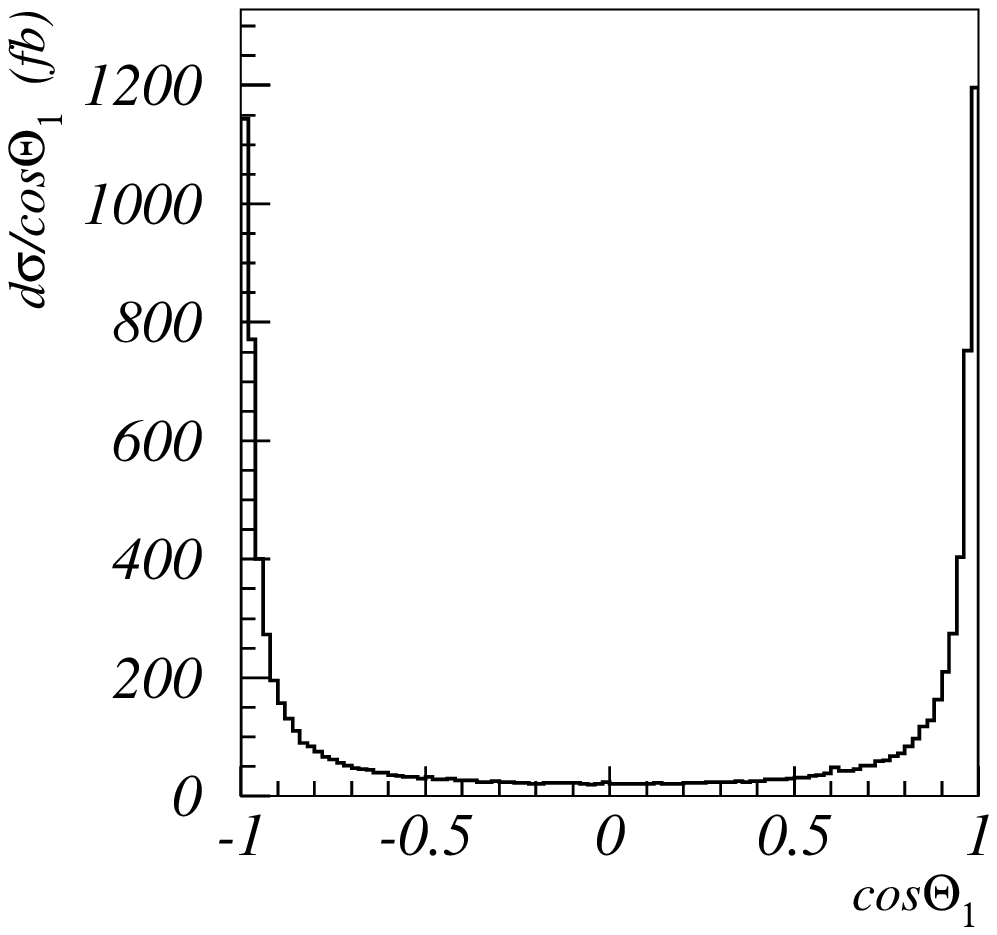}

 \vspace{-15pt}

{ Fig.3. The differential cross section  of $\gamma\gamma
\rightarrow e^+e^-\mu^+\mu^-$ process dependence on cosine of
polar angle  at averaged polarization state of interacting
particles.}

\end{figure}
 \end{center}

\vspace{-4.7cm}
\begin{figure}[ht!]
\leavevmode
\begin{minipage}[b]{.33\linewidth}
\centering
\includegraphics[width=\linewidth, height=7.5cm, angle=0]{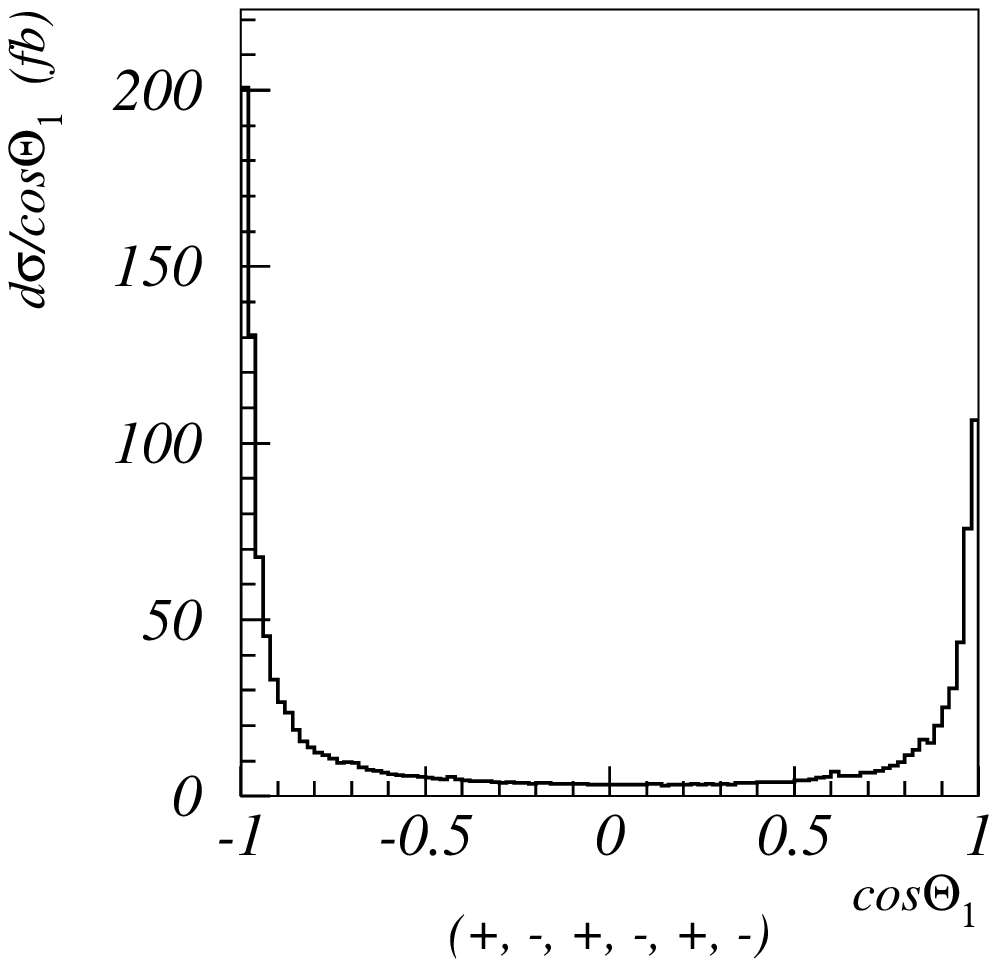}
\end{minipage}
\begin{minipage}[b]{.33\linewidth}
\centering
\includegraphics[width=\linewidth, height=7.5cm, angle=0]{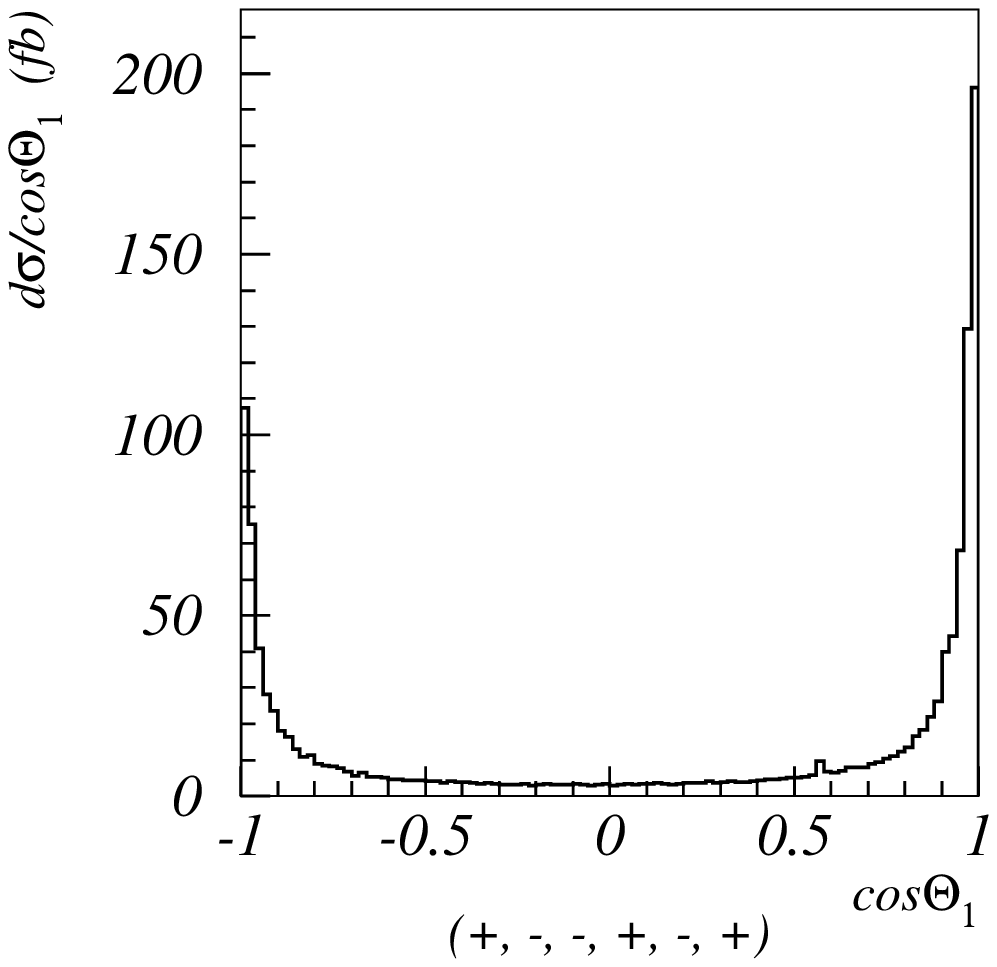}
\end{minipage}\hfill
\begin{minipage}[b]{.33\linewidth}
\centering
\includegraphics[width=\linewidth, height=7.5cm, angle=0]{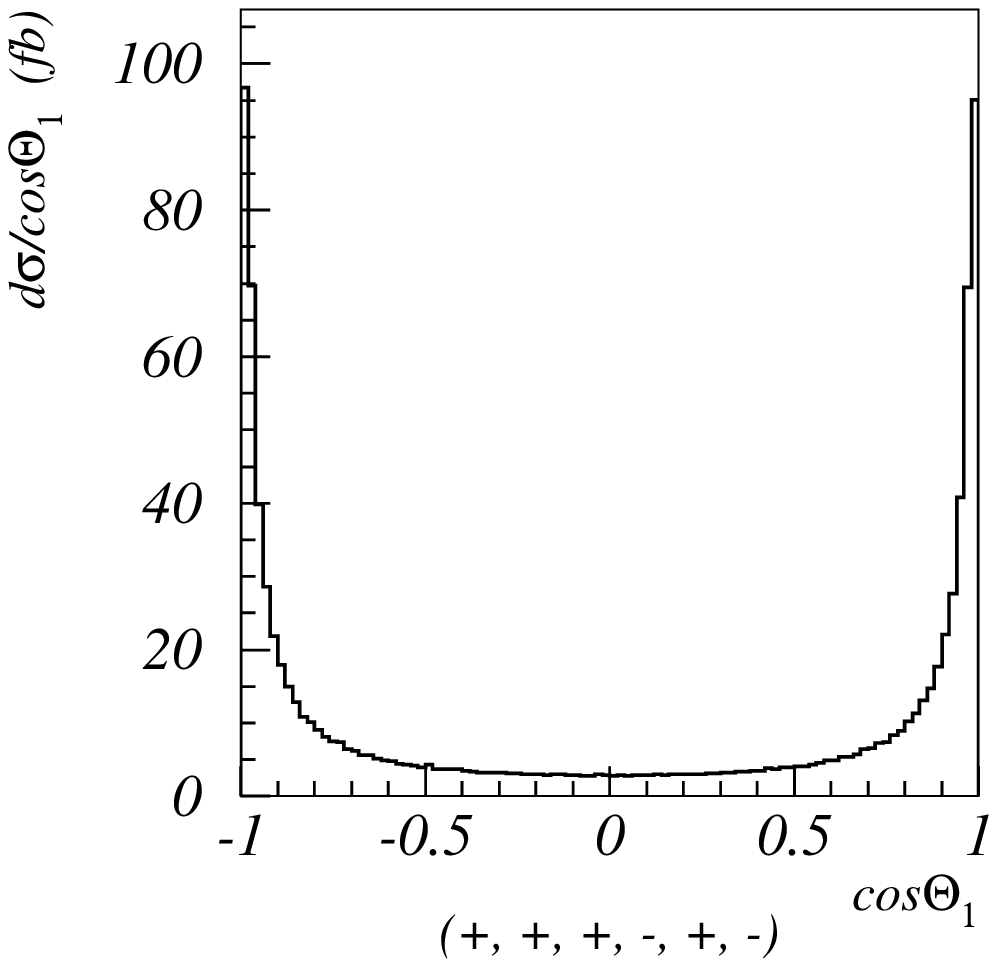}
\end{minipage}

 \vspace{-32pt}

 {\hspace{2.6cm}\tiny Fig.4. \hspace{4.9 cm} Fig.5. \hspace{4.9 cm} Fig.6.}
\end{figure}

\vspace{-0.7cm}

\vspace{-0.7cm}

\begin{figure}[ht!]
\leavevmode
\begin{minipage}[b]{.33\linewidth}
\centering
\includegraphics[width=\linewidth, height=7.5cm, angle=0]{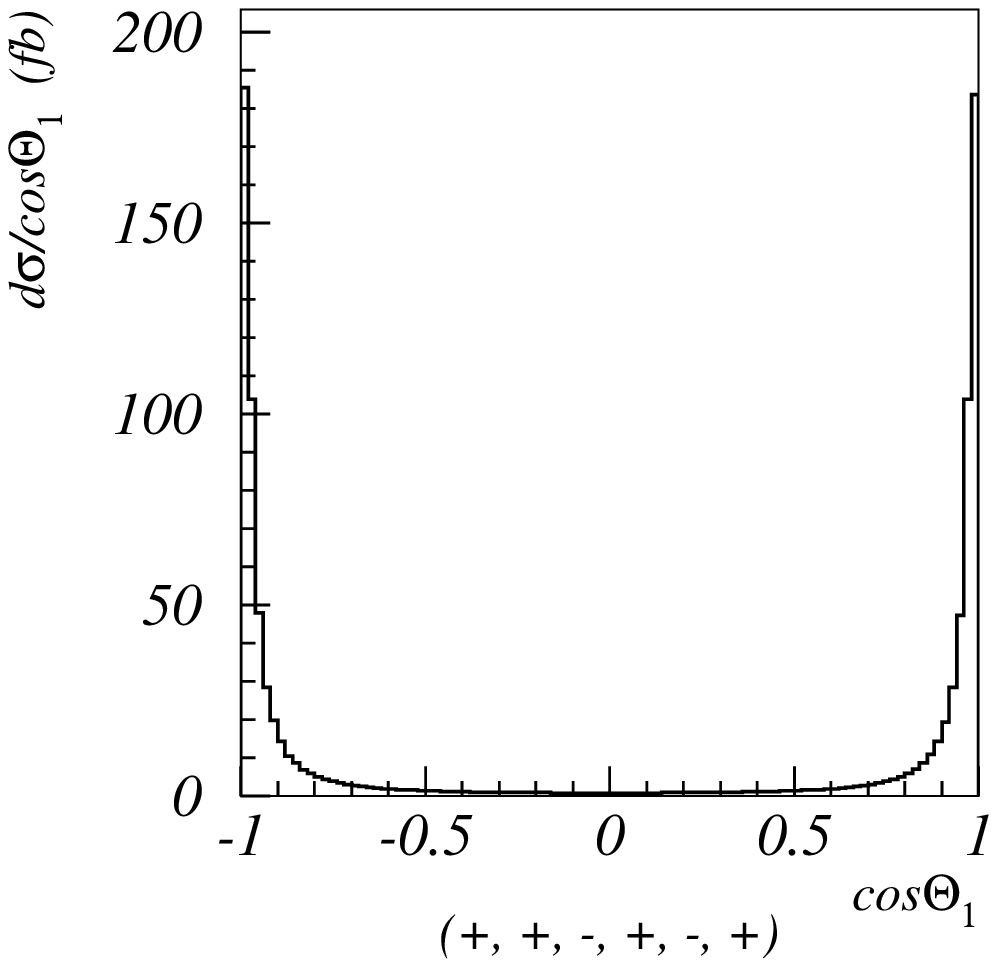}
\end{minipage}
\begin{minipage}[b]{.33\linewidth}
\centering
\includegraphics[width=\linewidth, height=7.5cm, angle=0]{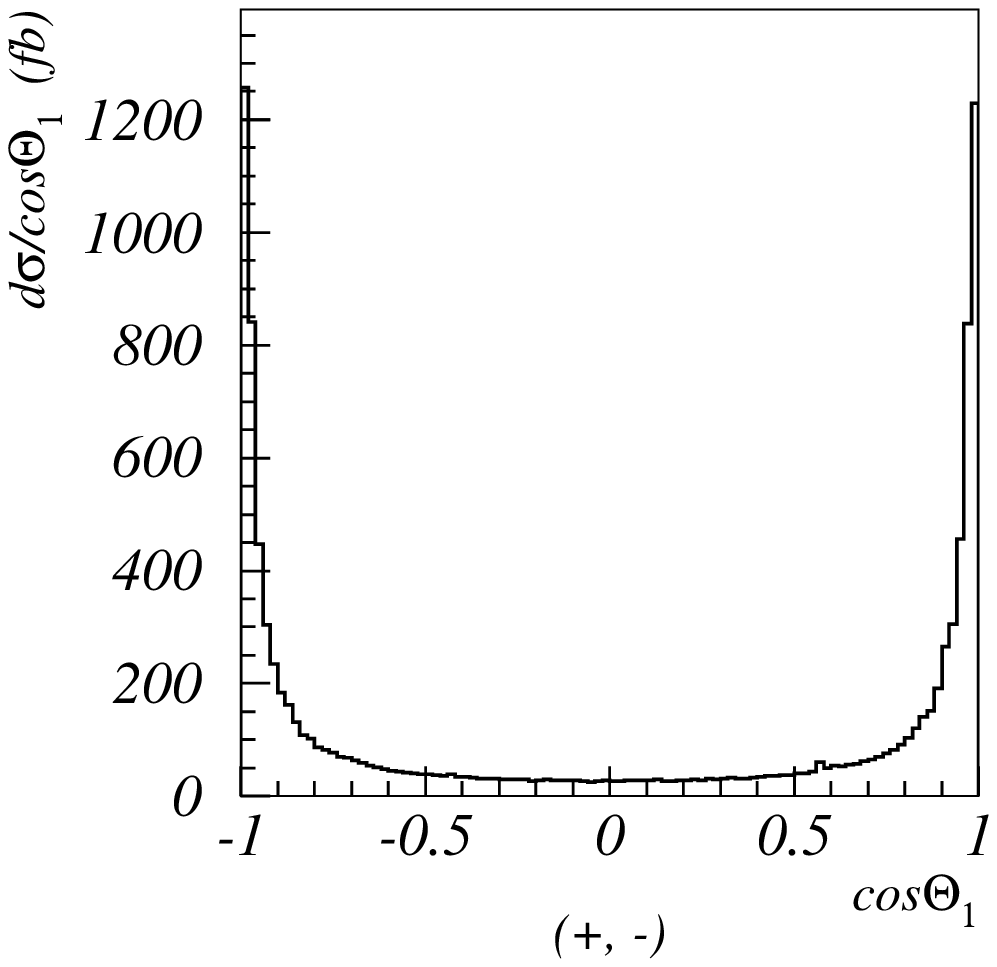}
\end{minipage}\hfill
\begin{minipage}[b]{.33\linewidth}
\centering
\includegraphics[width=\linewidth, height=7.5cm, angle=0]{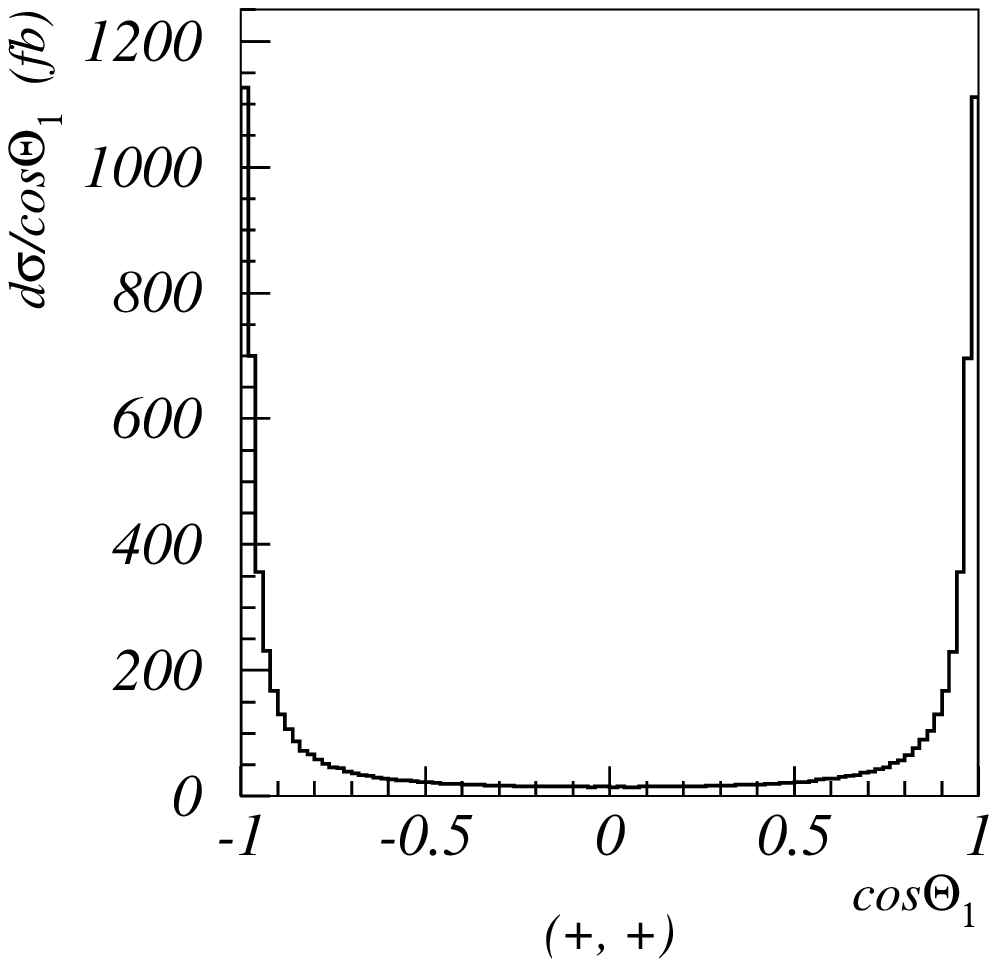}
\end{minipage}

 \vspace{-32pt}

 {\hspace{2.6 cm}\tiny Fig.7. \hspace{4.9 cm} Fig.8. \hspace{4.6 cm} Fig.9.}

\vspace{2cm}
\begin{center}

Figs.4-9. The  dependence  of the $\gamma\gamma \rightarrow
e^+e^-e^+e^-$ differential cross section on cosine of polar angle
at fixed spin states of interacting particles.  For identification
of particle polarization state the following notation is used:
$(+,-,+,-,+,-,)=(\lambda_1,\lambda_2,\lambda_3,\lambda_4,\lambda_5,\lambda_6)$,
where $\lambda_{1(2)}\;$ corresponds to polarization state of
photon with four momentum $k_{1(2)}$, $\lambda_{3,4,5,6}\;-$
helicity of lepton with four momentum $p_{1,2,3,4}$\;. Notations
of $(+,-)$ and $(+,+)$ correspond to cross sections of unpolarized
leptons production by scattering of circular polarized photon
beams . The energy of $\gamma\gamma$- beam is $ 120$ GeV in c.m.s.
\end{center}

\end{figure}

\begin{center}
\begin{figure}[ht!]
\leavevmode \centering
\includegraphics[width=8.5cm, height=7.5cm, angle=0]{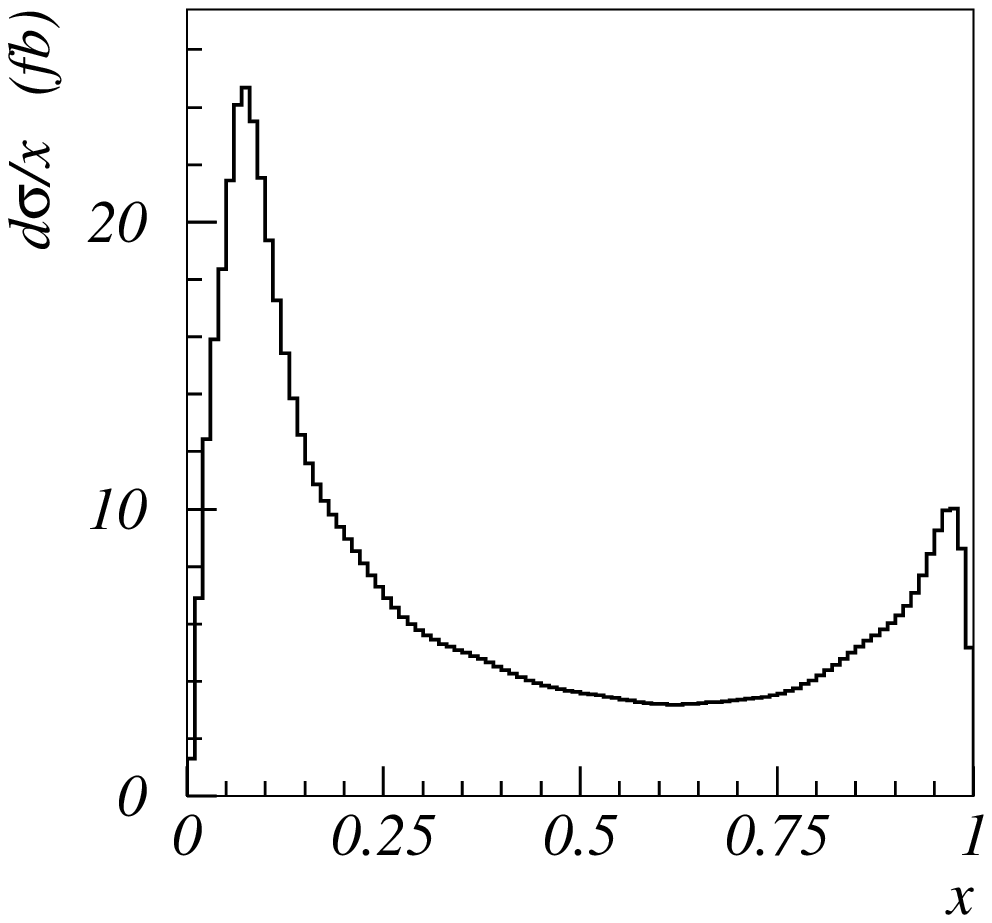}
\vspace{-15pt}

 Fig.10. The differential cross section  of
$\gamma\gamma \rightarrow e^+e^-e^+e^-$ process dependence on $x$.
Here $x=\frac{\displaystyle (e_1k_1)}{\displaystyle  s/2}$ is the
scalar product of four momenta of  initial photon and final
positron normalized by   invariant mass of  incoming photons pair.
\end{figure}
 \end{center}

\begin{center}
\begin{figure}[ht!]
\leavevmode \centering
\includegraphics[width=8.5cm, height=7.5cm, angle=0]{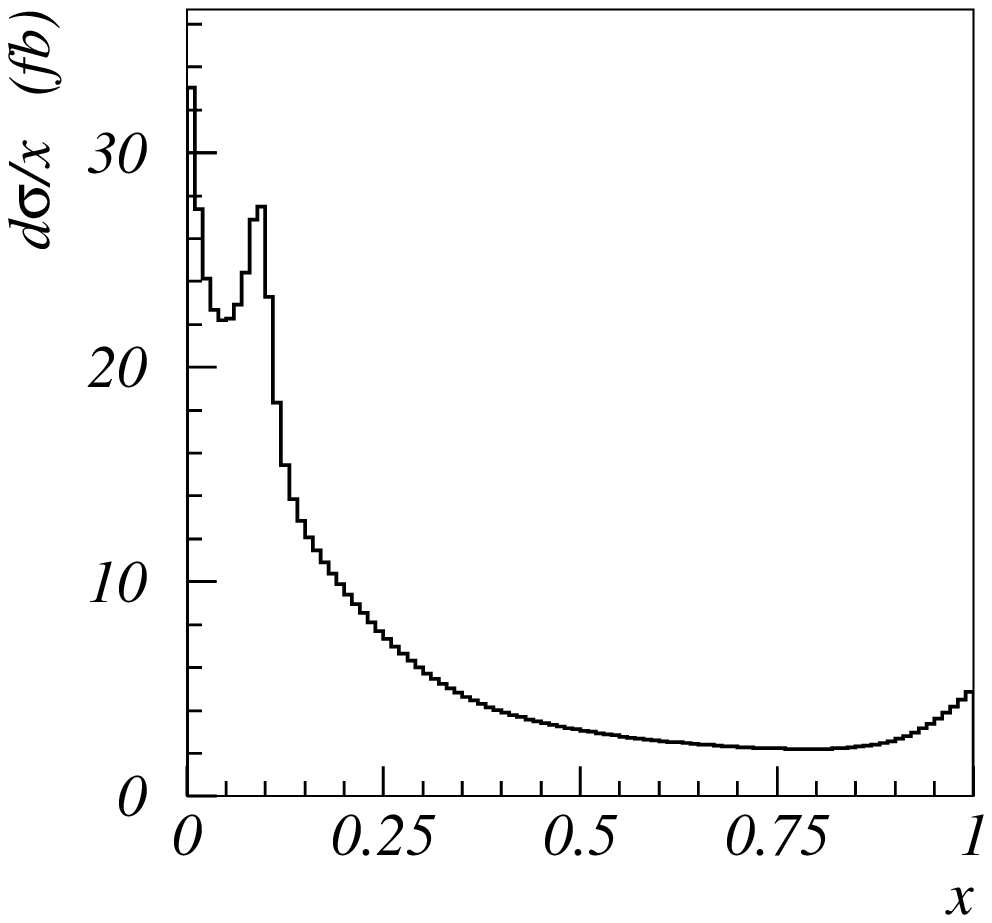}
\vspace{-15pt}

 Fig.11. The differential cross section  of
$\gamma\gamma \rightarrow e^+e^-e^+e^-$  process dependence on
$x$. Here $x=\frac{\displaystyle (e_1e_2)}{\displaystyle s/2}$ is
the scalar product of four momenta of  final electron and final
positron  normalized by   invariant mass of  incoming photons
pair.

\end{figure}
 \end{center}

\begin{center}
\begin{figure}[ht!]
\leavevmode \centering
\includegraphics[width=8.5cm, height=7.5cm, angle=0]{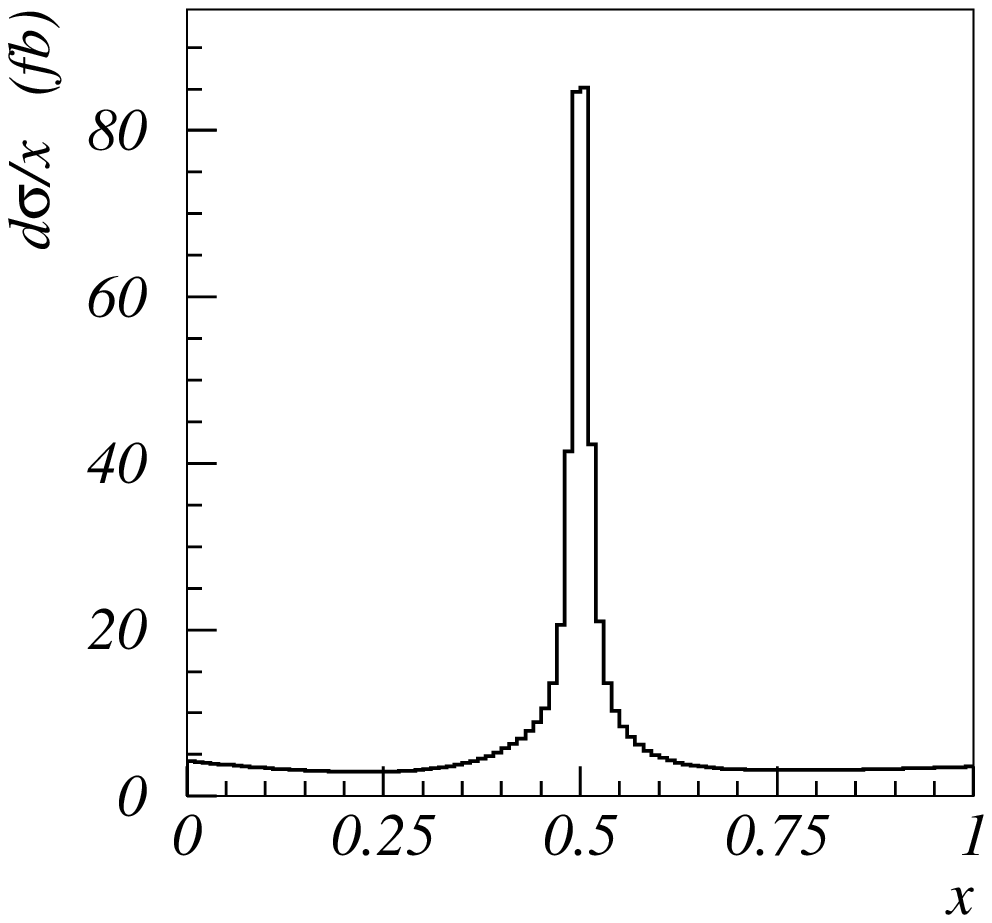}
\vspace{-15pt}

 Fig.12. The differential cross section  of
$\gamma\gamma \rightarrow e^+e^-e^+e^-$ process dependence  on
$x$. Here $x=\frac{\displaystyle E_1+E_2}{\displaystyle
\sqrt{s/2}}$ is the energy of any final electron-positron pair
normalized by energy of incoming photons pair.

\end{figure}
 \end{center}

\begin{center}
\begin{figure}[ht!]
\leavevmode \centering
\includegraphics[width=8.5cm, height=7.5cm, angle=0]{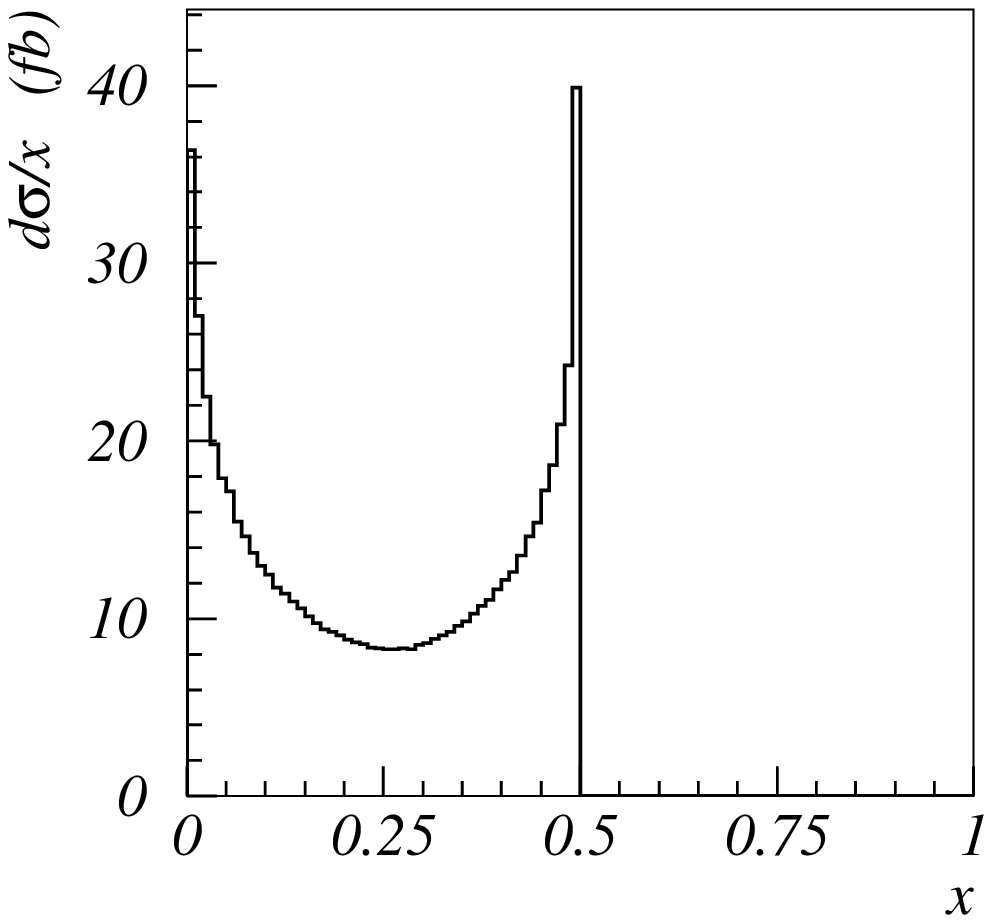}
\vspace{-15pt}

 Fig.13. The differential cross section  of
$\gamma\gamma \rightarrow e^+e^-e^+e^-$ process dependence on $x$.
Here $x=\frac{\displaystyle E_1}{\displaystyle \sqrt{s/2}}$ is the
energy of any final electron normalized by energy of incoming
photons pair.

\end{figure}
 \end{center}

\vspace{2cm}

\begin{figure}[ht!]
\leavevmode
\begin{minipage}[b]{1.\linewidth}
\centering
\includegraphics[width=\linewidth, height=10cm, angle=0]{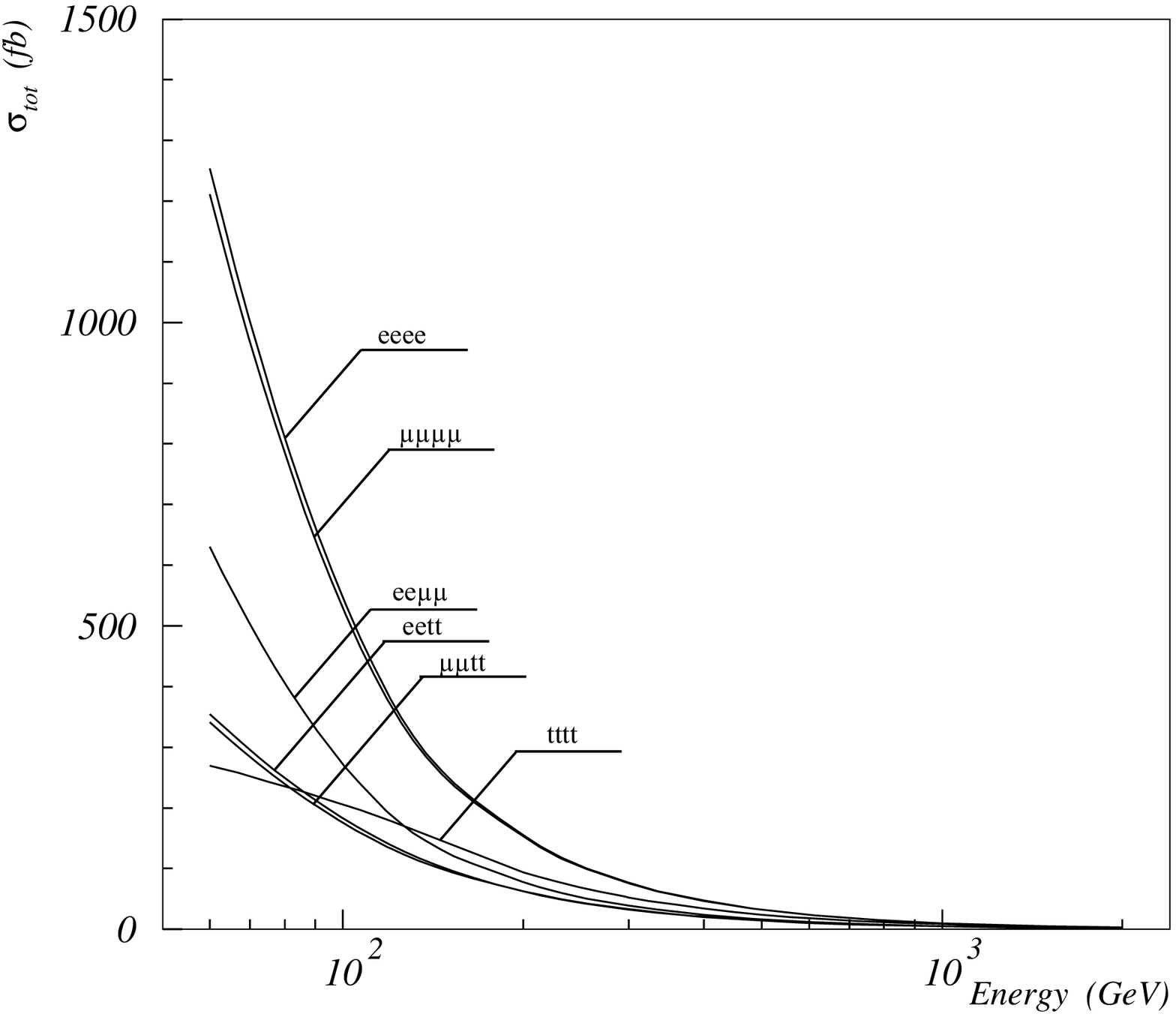}
\end{minipage}
 \vspace{-15pt}
\begin{center}
 {Fig.14. The total cross sections  of $\gamma\gamma\rightarrow 4l$ processes dependence on energy of interacting
 particles.}
\end{center}
\end{figure}

\begin{figure}[ht!]
\leavevmode
\begin{minipage}[b]{1.\linewidth}
\centering
\includegraphics[width=\linewidth, height=10cm, angle=0]{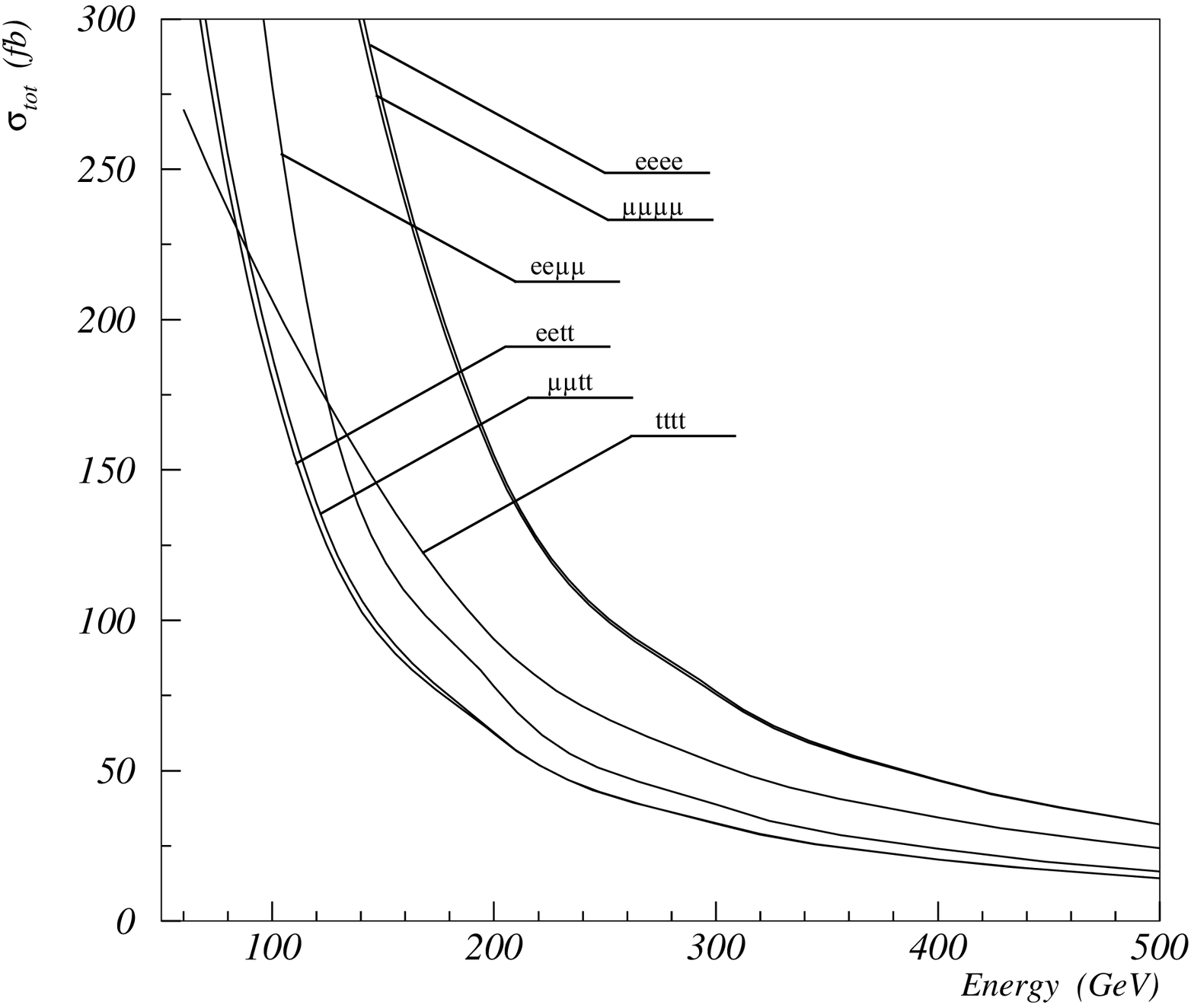}
\end{minipage}
 \vspace{-15pt}
\begin{center}
 {Fig.15. The total cross sections  of $\gamma\gamma\rightarrow 4l$ processes dependence on energy of interacting
 particles.}
\end{center}
\end{figure}

\newpage $$ $$\newpage $$ $$ \newpage \newpage \newpage \newpage \newpage \newpage $$
$$\newpage $$ $$\newpage $$ $$ $$ $$ $$ $$

\end{document}